\documentclass[a4paper]{jpconf}
\usepackage{graphicx}
\usepackage{bbm}
\usepackage{hyperref}
\usepackage{amsmath,amssymb,amsbsy}
\newcommand{\3}{\mbox{${\bf \underline{3}}$}}
\newcommand{\s}{\mbox{${\bf \underline{1}}$}}
\newcommand{\spr}{\mbox{${\bf \underline{1}'}$}}
\newcommand{\sppr}{\mbox{${\bf {\underline{1}''}}$}}
\begin{document}
\title{Suppression of flavor violation in an\\ A4 warped extra dimensional model}

\author{Avihay Kadosh }

\address{Center for Theoretical Physics, University of Groningen\\ 9747AG, Groningen, The Netherlands}

\ead{a.kadosh@rug.nl}

\begin{abstract}
In an attempt to simultaneously explain the observed masses and
mixing patterns of both quarks and leptons, we recently proposed a
model (JHEP08(2010)115) based on the non abelian discrete flavor
group A$_4$, implemented in a custodial RS setup with a bulk
Higgs. We showed that the standard model flavor structure can be
realized within the zero mode approximation (ZMA), with nearly TBM
neutrino mixing and a realistic CKM matrix with rather mild
assumptions. An important advantage of this framework with respect
to flavor anarchic models is the vanishing of the dangerous tree
level KK gluon contribution to $\epsilon_K$ and the suppression of
the new physics one loop contributions to the neutron EDM,
$\epsilon'/\epsilon$, $b\rightarrow s \gamma$ and Higgs mediated
flavor changing neutral curent (FCNC) processes. These results are
obtained beyond the ZMA, in order to account for the the full
flavor structure and mixing of the zero modes and first
Kaluza-Klein (KK) modes of all generations. The resulting
constraints on the KK mass scale are shown to be significantly
relaxed compared to the flavour anarchic case, showing explicitly
the role of non abelian discrete flavor symmetries in relaxing
flavor violation bounds within the RS setup. As a byproduct of our
analysis we also obtain the same contributions for the custodial
anarchic case with two $SU(2)_R$ doublets for each fermion
generation.
\end{abstract}

\section{Introduction}
Recently  we have proposed a model \cite{A4Warped}  based on a
bulk $A_4$ flavor symmetry \cite{a4} in warped geometry \cite{RS},
in an attempt to account for the  hierarchical charged fermion
masses, the hierarchical mixing pattern in the quark sector and
the large mixing angles and the mild hierarchy of masses in the
neutrino sector. In analogy with a previous RS realization of
A$_{4}$ for the lepton sector \cite{Csaki:2008qq}, the three
generations of left-handed quark doublets are unified into a
triplet of $A_4$; this assignment forbids tree level  FCNCs driven
by the exchange of KK gauge bosons. The scalar sector of the
RS-A${}_4$ model consists of two bulk flavon fields, in addition
to a bulk Higgs field. The bulk flavons transform as triplets of
$A_{4}$, and allow for a complete
 "cross-talk" \cite{Volkas} between the $A_{4}\to Z_{2}$
spontaneous symmetry breaking (SSB) pattern associated with the
heavy neutrino sector - with scalar mediator  peaked towards the
UV brane - and the $A_{4}\to Z_{3}$ SSB pattern associated with
the quark and charged lepton sectors - with scalar mediator peaked
towards the IR brane - and allows to obtain realistic masses and
almost realistic mixing angles in the quark sector. A bulk
custodial symmetry, broken differently at the two branes
\cite{Agashe:2003zs}, guarantees the suppression of large
contributions to electroweak precision observables
\cite{Carena:2007}, such as the Peskin-Takeuchi $S$, $T$
parameters. However, the mixing  between zero modes of the 5D
theory and their Kaluza-Klein (KK) excitations -- after 4D
reduction -- may still cause significant new physics (NP)
contributions to SM suppressed flavor changing neutral current
(FCNC) processes.

\noindent In general, when no additional flavor symmetries are
present and the 5D Yukawa matrices are anarchical, FCNC processes
are already generated at the tree level  by a KK gauge boson
exchange \cite{Agashe:2004cp}. Stringent constraints on the KK
scale come from the $K^{0}-\overline{K^{0}}$ oscillation
 parameter $\epsilon_{K}$, the radiative decays $b\to s(d)\gamma$ \cite{Agashe:2004cp,Azatov},
the direct CP violation parameter $\epsilon^\prime/\epsilon_K$
\cite{IsidoriPLB}, and especially the neutron electric dipole
moment (EDM) \cite{Agashe:2004cp}, also in the presence of an
RS-GIM suppression mechanism \cite{rsgim1,Cacciapaglia:2007fw}.

\noindent Conclusions may differ if a flavor pattern of the Yukawa
couplings is assumed to hold in the 5D theory due to bulk flavor
symmetries. They typically imply an increased alignment between
the 4D fermion mass matrix and the Yukawa and gauge couplings,
thus suppressing the amount of flavor violation induced by the
interactions with KK states.

\noindent In our case, the most relevant consequence of imposing
an $A_4$ flavor symmetry is  the degeneracy of the left-handed
fermion bulk profiles $f_Q$, i.e. $diag(f_{Q_1,Q_2,Q_3})=f_Q\times
\mathbbm{1}$. In addition, the distribution of phases, CKM and
Majorana-like, in the mixing matrices might induce zeros in the
imaginary components of the Wilson coefficients contributing to CP
violating quantities.

\noindent In the following we review the most relevant features of
the RS-A$_4$ model \cite{A4Warped} and its phenomenological
consequences \cite{A4CPV}. In particular, we discuss helicity
flipping FCNC processes induced by dipole operators and obtain the
most significant constraints coming from $b\rightarrow s\gamma$,
${\mbox Re}(\epsilon'/\epsilon_K)$ and the neutron EDM. As a
byproduct we obtain the analogous constraints in a custodial
anarchic scheme with a bulk Higgs. Implications of adopting a
$P_{LR}$ extended custodial setup \cite{AgashePLR} are then
briefly discussed.

\section{The RS-A$_4$ model}
\begin{figure}
\centering
\includegraphics[width=12truecm]{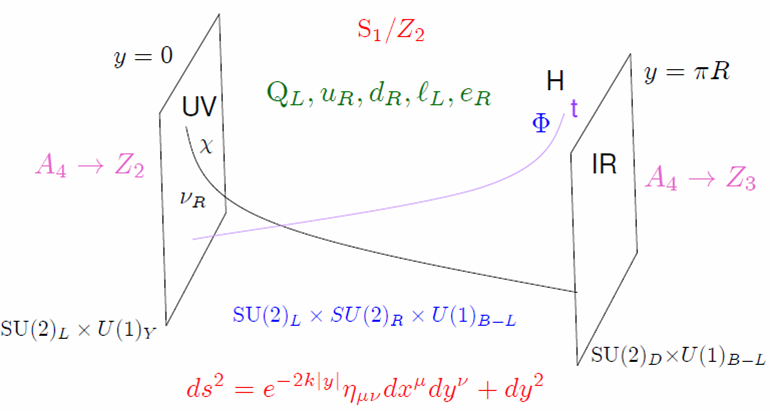}
\caption{A pictorial description of the RS-A$_4$ setup. The bulk
geometry is described by the metric at the bottom and $k\simeq
M_{Pl}$ is the $AdS_5$ curvature scale. All fields propagate in
the bulk and the UV(IR) peaked nature of the heavy RH neutrinos,
the Higgs field, the $t$ quark and the A$_4$ flavons, $\Phi$ and
$\chi$, is emphasized. The SSB patterns of the bulk symmetries on
the UV and IR branes are specified on the side (for A$_4$) and on
the bottom (for $SU(2)_L\times SU(2)_R\times U(1)_{B-L}$) of each
brane. }\label{ModelScheme}
\end{figure}
 The RS-A$_4$ setup  \cite{A4Warped} is illustrated  in
Fig.~\ref{ModelScheme}. The bulk geometry is that of a slice of
$AdS_5$ compactified on an orbifold $S_1/Z_2$ \cite{RS} and is
described by the metric on the bottom of Fig.~\ref{ModelScheme}.
All 5D fermionic fields propagate in the bulk and transform under
the following representations of $\left(SU(3)_c\times
SU(2)_L\times SU(2)_R\times U(1)_{B-L}\right)\times
A_4$\cite{A4Warped}:
\begin{equation}
\begin{array}{c}
Q_L \sim \left( 3,2,1,\frac{1}{3} \right) \left( \3 \right) \\
\\
u_R \oplus u'_R \oplus u''_R \sim \left( 3,1,2,\frac{1}{3}
\right)\left(\s \oplus
\spr \oplus \sppr \right) \\
\\
d_R \oplus d'_R \oplus d''_R \sim \left( 3,1,2,\frac{1}{3}
\right)\left(\s \oplus \spr \oplus \sppr \right)
\end{array}
\quad\
\begin{array}{c}
\ell_L \sim \left( 1,2,1,-1 \right) \left( \3 \right) \\
\\
\nu_R \sim \left( 1,1,2,0 \right)\left( \3 \right) \\
\\
e_R \oplus e'_R \oplus e''_R \sim \left( 1,1,2,-1 \right)\left(\s
\oplus \spr \oplus \sppr \right)\, .
\end{array}
\end{equation}
The SM fermions (including RH neutrinos) are identified with the
zero modes of the 5D fermions above. The zero (and KK) mode
profiles are determined by the bulk mass of the corresponding 5D
fermion, denoted by $c_{q_L,u_i,d_i}k$ (and BC)\cite{A4CPV}. The
scalar sector contains the IR peaked Higgs field and the UV and IR
peaked flavons, $\chi$ and $\Phi$, respectively. They transform
as:
\begin{equation}
\Phi \sim \left( 1,1,1,0 \right) \left( \3 \right),\quad \chi \sim
\left( 1,1,1,0 \right) \left( \3 \right),\quad H\left(1,2,2,0
\right) \left( \s \right)\, .
\end{equation}
The SM Higgs field is identified with the first KK mode of $H$.
All fermionic zero modes acquire masses through Yukawa
interactions with the Higgs field and the $A_4$ flavons after SSB.
The 5D $\left(SU(3)_c\times SU(2)_L\times SU(2)_R\times
U(1)_{B-L}\right)\times A_4$ invariant Yukawa Lagrangian will
consist of leading order(LO) UV/IR peaked interactions and next to
leading order (NLO) "cross-talk" and "cross-brane"
interactions\cite{A4Warped}.
 The
LO interactions in the neutrino sector are shown in
\cite{A4Warped} using the see-saw I mechanism, to induce a
tribimaximal (TBM)\cite{TBM} pattern for neutrino mixing while NLO
"cross brane" and "cross talk" interactions, induce small
deviations of $\mathcal{O}(0.04)$, which are still in good
agreement with the current experimental bounds \cite{NuFogli}.
Here, we focus on the quark sector and on the phenomenology
relevant for flavor and CP violating processes. The relevant terms
of the 5D Yukawa lagrangian are of the following form:

\begin{equation}
\mathcal{L}_{5D}^{Y\!uk.}=\mathcal{L}_{LO}+\mathcal{L}_{NLO}=\frac{1}{k^2}\overline{Q}_L\Phi
H
(u_R^{(\prime,\,\prime\prime)},d_R^{(\prime,\,\prime\prime)})+\frac{1}{k^{3/2}}\overline{Q}_L\Phi\chi
H(u_R^{(\prime,\,\prime\prime)},d_R^{(\prime,\,\prime\prime)})\label{LYuk5D}\end{equation}

\noindent Notice that the LO interactions are peaked towards the
IR brane while the NLO interactions mediate between the two branes
due to the presence of both $\Phi$ and $\chi$.

 \noindent The VEV and physical profiles
for the bulk scalars are obtained by solving the corresponding
equations of motion with a UV/IR localized quartic potential term
and an IR/UV localized  mass term \cite{WiseScalar}. In this way
one can obtain either UV or IR peaked and also flat profiles
depending on the bulk mass and the choice of boundary conditions.
The resulting VEV profiles of the RS-A$_4$ scalar sector are:
\begin{equation}
v_{H(\Phi)}^{5D}=H_0(\phi_0)e^{(2+\beta_{H(\phi)})k(|y|-\pi
R)}\qquad
v_\chi^{5D}=\chi_0e^{(2-\beta_\chi)k|y|}(1-e^{(2\beta_\chi)k(|y|-\pi
R)})\,,\label{VEVprofile}
\end{equation}
where $\beta_{H\Phi,\chi}=\sqrt{4+\mu_{H,\Phi,\chi}^2}$, and
$\mu_{H,\Phi,\chi}$ is the bulk mass of the corresponding scalar
in units of $k$, the cutoff of the 5D theory. The following vacua
for the Higgs and the $A_4$ flavons $\Phi$ and $\chi$
\begin{equation}
\langle
\Phi\rangle=(v_\phi,v_\phi,v_\phi)\qquad\langle\chi\rangle=(0,v_\chi,0)\qquad
\langle H\rangle=v_{\footnotesize
H}\left(\begin{array}{cc}1&0\\0&1\end{array}\right)\label{VEValignment},\end{equation}
provide at LO TBM neutrino mixing and zero quark mixing
\cite{a4,Volkas}. The stability of the above vacuum alignment is
discussed in \cite{A4Warped}. The VEV of $\Phi$ induces an A$_4\to
Z_3$ SSB pattern, which in turn induces no quark mixing and is
peaked towards the IR brane. Similarly, the VEV of $\chi$ induces
an A$_4\to Z_2$ SSB pattern peaked towards the UV brane and is in
charge of the TBM mixing pattern in the neutrino sector.
Subsequently, NLO interactions break A$_4$ completely and induce
quark mixing and deviations from TBM, both in good agreement with
experimental data. The Higgs VEV is in charge of the SSB pattern
$SU(2)_L\times SU(2)_R\rightarrow SU(2)_D$, which is peaked
towards the IR brane. The (gauge) SSB pattern on the UV brane is
driven by orbifold BC and a planckian UV localized VEV, which is
effectively decoupled from the model.

\noindent To summarize the implications of the NLO interactions in
the quark sector, we provide the structure of the LO+NLO quark
mass matrices in the ZMA \cite{A4Warped}:
\begin{equation} \frac{1}{v}(M+\Delta M)_{u,d}=\underbrace{\left(\begin{array}{ccc}y_{u,d}^{4D}&
 y_{c,s}^{4D} & y_{t,b}^{4D}
\\y_{u,d}^{4D}& \omega y_{c,s}^{4D} & \omega^2y_{t,b}^{4D}\\y_{u,d}^{4D}& \omega^2y_{c,s}^{4D} & \omega
y_{t,b}^{4D}\end{array}\right)}_{\sqrt{3}U(\omega)
diag(y_{u_i,d_i}^{4D})}+\left(\begin{array}{ccc}
f_\chi^{u,d}\tilde{x}_1^{u,d}& f_\chi^{c,s}\tilde{x}_2^{u,d}&
f_\chi^{t,b}\tilde{x}_3^{u,d}\\0&0&0\\f_\chi^{u,d}\tilde{y}_1^{u,d}&
f_\chi^{c,s}\tilde{y}_2^{u,d}&
f_\chi^{t,b}\tilde{y}_3^{u,d}\end{array}\right)\label{MDeltaM}
\end{equation}
where $\omega=e^{2\pi i/3}$, $v=174$GeV is the 4D Higgs VEV,
$y^{4D}_{u,c,t,d,s,b}$ are the effective 4D LO  Yukawa couplings
and $\tilde{x}_i^{u,d}$, $\tilde{y}_i^{u,d}$ are the coefficients
of the 5D NLO Yukawa interactions. The function
$f_\chi^{u_i,d_i}\simeq 2\beta_\chi
C_\chi/(12-c_{q_L}-c_{u_i,d_i})\simeq0.05$ is the characteristic
suppression of the 4D effective NLO Yukawa interactions and
$C_\chi=\chi_0/M_{Pl}^{3/2}\simeq0.155$. Finally, the unitary
matrix, $U(\omega)$ is the LO left diagonalization matrix in both
the up and down sectors, $(V_L^{u,d})_{LO}$, which is independent
of the LO Yukawa couplings, while
$(V_R^{u,d})_{LO}=\mathbbm{1}$(see \cite{A4Warped}). Using
standard perturbative techniques on the matrix in
Eq.~(\ref{MDeltaM}) we obtained $(V_{L,R}^{u,d})_{NLO}$
\cite{A4Warped,A4CPV} at
$\mathcal{O}\Big(f_\chi^{u_i,d_i}(\tilde{x}_i^{u,d},\tilde{y}_i^{u,d})\Big)$
and showed that by a rather mild deviation from a universality
assumption on the magnitudes and phases of the NLO coefficients,
$\tilde{x}_i^{u,d},\tilde{y}_i^{u,d}$, we are able to obtain  an
almost realistic CKM matrix of the characteristic form
\cite{A4Warped}:
\begin{equation} V_{CKM}=\left(\begin{array}{ccc}1 &
 a\lambda & b\lambda^3\\-a^*\lambda & 1 & c\lambda^2\\
-b^*\lambda^3 & -c^*\lambda^2 & 1
\end{array}\right).\end{equation}


A precise matching of the CKM matrix including deviations from
unity of diagonal elements, $|V_{ub}|\neq |V_{td}|$ and phase
structure has to be performed at higher order in
$f_\chi^{u_i,d_i}(\tilde{x}_i^{u,d},\tilde{y}_i^{u,d})$.

\section{Phenomenology of RS-A$_4$ and constraints on the KK scale}
The main difference between the RS-A$_4$ setup and an anarchic RS
flavor scheme \cite{Agashe:2004cp} lies in the degeneracy of
fermionic LH bulk mass parameters, which implies the universality
of LH zero mode profiles and hence forbids gauge mediated FCNC
processes at tree level, including the KK gluon exchange
contribution to $\epsilon_K$. The latter provides the most
stringent constraint on flavor anarchic models, together with the
neutron EDM \cite{Agashe:2004cp,IsidoriPLB}. However, the choice
of the common LH bulk mass parameter, $c_q^L$ is strongly
constrained by the matching of the top quark mass
($m_t(1.8$\,TeV)$\approx140$\,TeV) and the perturbativity bound of
the 5D top Yukawa coupling, $y_t$. Most importantly, when
considering the tree level corrections to the $Zb\bar{b}$ coupling
 against the stringent EWPM at the Z pole, we
realize \cite{A4Warped} that for an IR scale,
$\Lambda_{IR}\simeq1.8\,$TeV and $m_h\approx 200$\,GeV, $c_q^L$ is
constrained to be larger than 0.35. Assigning $c_q^L=0.4$ and
matching with $m_t$ we obtain $y_t<3$, which easily satisfies the
5D Yukawa perturbativity bound. The constraint on $c_q^L$ from
$Zb\bar{b}$ has a moderate dependence on the Higgs mass, such that
the constraint $\Lambda_{IR}>1.8$\,TeV for $c_q^L=0.35$ and
$m_h\approx200 GeV$, is relaxed to $\Lambda_{IR}>1.3$\,TeV for
$m_h\approx 1$TeV \cite{A4CPV}.

\section{Dipole Operators and FCNC processes}

FCNC processes provide us with some of the most stringent
constraints for physics beyond the SM and this is also the case
for the framework of flavor anarchic models in warped extra
dimensions \cite{Agashe:2004cp,Azatov,IsidoriPLB}. In the quark
sector, significant bounds on the KK mass scale typically arise
from the neutron electric dipole moment (EDM), the CP violation
parameters, $Re(\epsilon^\prime/\epsilon_K)$ and $\epsilon_K$, and
radiative $B$ decays such as $b\to s\gamma$. All these processes
are mediated by effective dipole operators. It is also well known
\cite{Buras} that SM interactions only induce, to leading order,
the single chirality operators $O_{7\gamma}$ and $O_{8g}$ $(j>i)$
\begin{equation}
O_{7\gamma(8g)}=\bar{d}_L^i \sigma^{\mu\nu}d_R^j
F_{\mu\nu}(G_{\mu\nu}).\label{SM-O7}\end{equation}

\begin{figure}
\begin{minipage}[t]{0.5\linewidth}
\begin{center}
\includegraphics[scale=0.60]{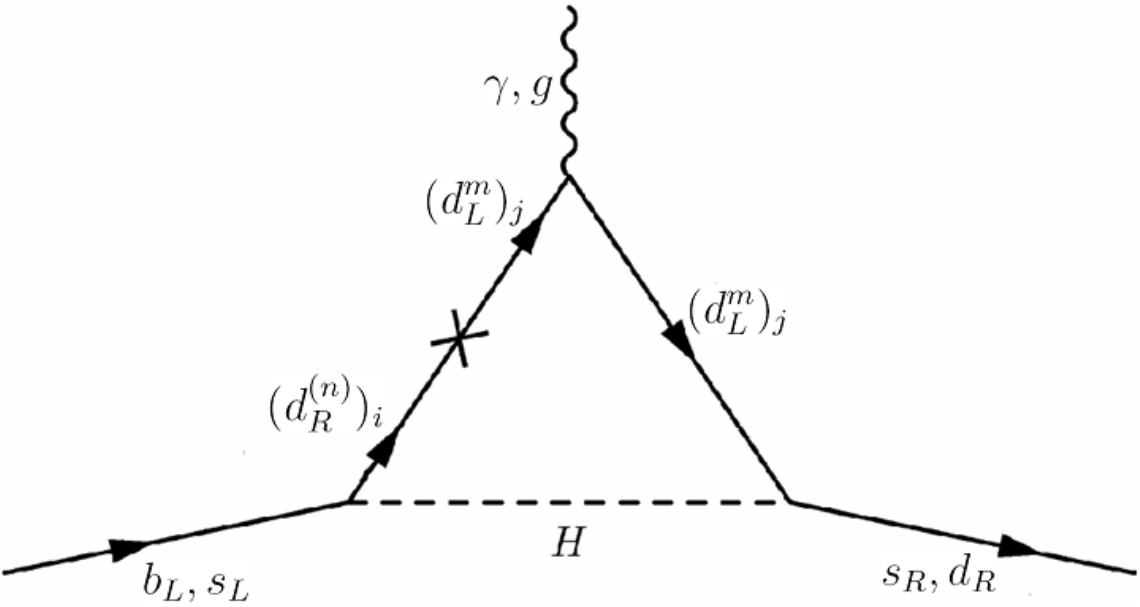}
\caption{One-loop down-type (neutral Higgs) contribution to
$b\rightarrow s\gamma$, $\epsilon'/\epsilon_K$ and the neutron EDM
(for external $d$ quarks). The analogous one-loop up-type
contribution (charged Higgs) contains internal up-type KK modes. }
\label{fcncLoop}
\end{center}
\end{minipage}
\hspace{0.2cm}
\begin{minipage}[t]{0.5\linewidth}
\begin{center}
\includegraphics[scale=0.26]{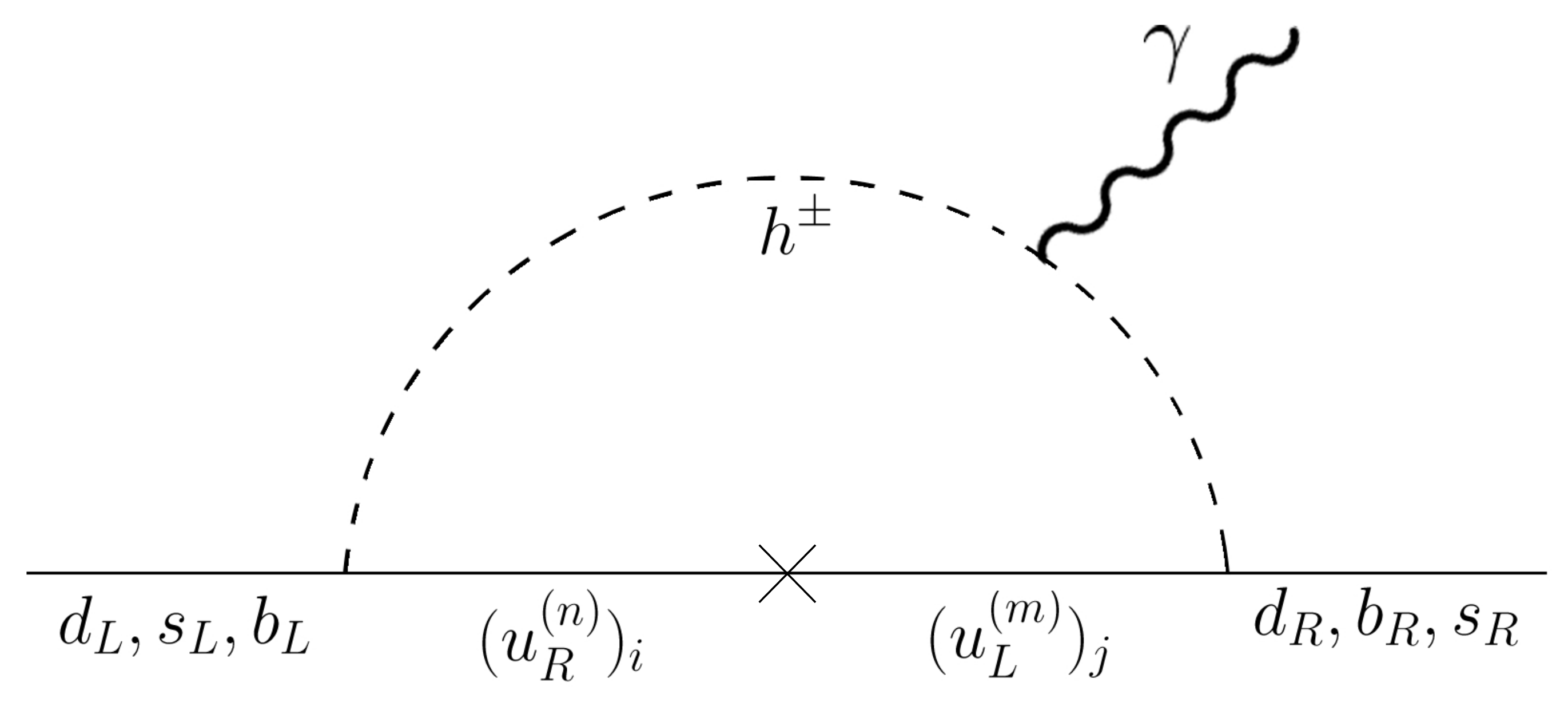}
\caption{Charged Higgs one-loop contribution to $b\rightarrow
s\gamma$ and  the neutron EDM. The latter has external $d$ quarks.
} \label{fcncLoopcharged}
\end{center}
\end{minipage}
\end{figure}

\noindent The new RS-A$_4$  contributions  to the  FCNC processes
we are interested in are generated at one-loop by the Yukawa
interactions between SM fermions  and their KK excitations,
leading to the diagrams shown in Fig.~\ref{fcncLoop}  and
Fig.~\ref{fcncLoopcharged}. 

In the mass insertion approximation, the flavor part of the
corresponding amplitude can be written in terms of spurions
\cite{Agashe:2004cp}. To illustrate it, we focus on the down type
contribution to the dipole operators in Fig.\ref{fcncLoop}, for
which the corresponding Wilson coefficient is given by:

\begin{equation}
(C_{7\gamma(8g)}^{d-type})_{ij}=A^{1L}\frac{v}{M_{KK}}\,\left(F_Q\,
\hat{Y}_d\hat{Y}_d^\dagger
\hat{Y}_d\,F_d\right)_{ij}\,,\label{C7DY}\end{equation} where
 $\hat{Y}_d$ are the 5D Yukawa matrices and the fermion profile matrices are
$F_{Q,u,d}=diag(f_{Q_i,u_j,d_j}^{-1})=(1/\sqrt{2k})diag(\hat{\chi}_{0_{Q_i,u_i,d_i}})$
and $\hat{\chi}_{0_{Q_i,u_i,d_i}}$ is the canonically normalized
zero mode wave function evaluated on the IR brane. Finally, the
factor $A^{1L}= 1/(64\pi^2M_{KK})$ comes from the one-loop
integral for the diagram in Fig.~\ref{fcncLoop}.

To account for overlap effects and illustrate the flavor patterns
generated in RS-A$_4$ we write the LO  mass matrix for the first
generation in the down-type sector following
\cite{Azatov,IsidoriPLB}, including the zero modes and first level
KK modes and overlap effects \cite{A4CPV}
\begin{equation}\frac{\hat{\mathbf{M}}_d^{KK}}{(M_{KK})}=
\left(\begin{array}{c}
\bar{Q}_L^{d(0)}\\\bar{d}_L^{(1^{--})} \\ \bar{Q}^{d(1)}_L\\
\bar{\tilde{d}}_L^{(1^{+-})} \end{array} \right)^T \left(
\begin{array}{cccc}
\breve{y}_df^{-1}_Q f^{-1}_d r_{00} x & 0 & \breve{y}_df^{-1}_Q
r_{01} x &
\breve{y}_uf^{-1}_Q r_{101} x \\
 0 &  \breve{y}_d^*r_{22} x & 1 & 0 \\
 \breve{y}_df^{-1}_d r_{10} x & 1 &  \breve{y}_dr_{11} x &  \breve{y}_ur_{111} x \\
 0 &  \breve{y}_u^*r_{222} x & 0 & 1
\end{array}
\right)\left(\begin{array}{c}
d_R^{(0)}\\Q_R^{d(1^{--})} \\ d^{(1)}_R\\
\tilde{d}_R^{(1^{-+})} \end{array} \right),\label{M4KK}
\end{equation}
where we factorized a common KK mass scale $M_{KK}$,
$\breve{y}_{u,d}\equiv 2y_{u,d}v_\Phi^{4D}e^{k\pi R}/k$ and the
perturbative expansion parameter is defined as $x\equiv v/M_{KK}$.
In the above equation the various $r$'s denote the ratio of the
bulk and IR localized effective couplings of the modes
corresponding to the matrix element in question. For simplicity,
we define $r_{111}\equiv r_{11^{-+}}$, $r_{101}\equiv
r_{01^{-+}}$, $r_{22}\equiv r_{1^-1^-}$, $r_{222}\equiv
r_{1^{-}1^{+-}}$ and the notation for the rest of the overlaps is
straightforward. The corresponding Yukawa matrix,
$\hat{Y}^{d}_{KK}$  is obtained by simply eliminating $x$ and the
1's from the above matrix and it leads the  flavor structure of
the contributions of $(++)$ and $(-+)$ KK modes to the processes
in Figs.~\ref{fcncLoop} and \ref{fcncLoopcharged}, including
overlap effects \cite{A4CPV},
\begin{eqnarray}
(C_{7\gamma(8g)}^{d,u})_{(++)}&\propto&
F_Q\hat{Y}_{d,u}\,r_{01}(c_{Q_i},c_{d_{\ell_1},u_{\ell_1}},\beta)\,\hat{Y}_{d,u}^\dagger\,r_{11}
(c_{d_{\ell_1},u_{\ell_1}},c_{Q_{\ell_2}},\beta)
\,\hat{Y}_{d,d}\,r_{10}(c_{Q_{\ell_2}},c_{d_j,d_j})F_{d}\nonumber\,,\\
(C_{7\gamma(8g)}^{d,u})_{(-+)}  &\propto&
F_Q\hat{Y}_{u,d}\,r_{01^{-+}}(c_{Q_i},c_{u_{\ell_1},d_{\ell_1}},\beta)\,\hat{Y}_{u,d}^\dagger\,
r_{1^{-+}1}(c_{u_{\ell_1},d_{\ell_1}},c_{Q_{\ell_2}},\beta)\,
\hat{Y}_{d,d}\,r_{10}(c_{Q_{\ell_2}},c_{d_j,d_j},\beta)F_{d}\,.\nonumber\\
\label{Spurion+Overlap}\end{eqnarray} Considering the degeneracy
of LH bulk mass parameters in the RS-A$_4$ setup and using the ZMA
diagonalization matrices to rotate the corresponding contribution
to the mass basis we obtain for the down type contribution in
Fig.~\ref{fcncLoop} \cite{A4CPV}:

\begin{eqnarray} (C_{7\gamma(8g)}^{d-type})_{ij}&=&
\frac{m_{d_i}A^{1L}f_Q^2}{v^2M_{KK}} \left [V_R^{d\dagger}{\mbox
diag}(f^2_{d,s,b})(\hat{r}_{00}^d)^{-1}
\tilde{r}_{01}^d\,\tilde{r}_{11}^d(\hat{r}_{00}^d)^{-1}
V_R^d\,{\mbox diag} (m^2_{d,s,b})\right.\nonumber\\&&\left. \times
V_R^{d\dagger}\,(\hat{r}_{00}^d)^{-1}\hat{r}_{10}^d\,
 V_R^{d\phantom{\dagger}}\right]_{ij}
 ,\label{C7Doverlap}
\end{eqnarray}
where
$\tilde{r}_{01}^{u,d}\tilde{r}_{11}^{u,d}=\hat{r}_{01}^{u,d}\hat{r}_{11}^{u,d}+
\hat{r}_{01^{-+}}^{u,d}\hat{r}_{1^{-+}1}^{u,d}$ and all overlap
matrices are real and diagonal. Since the overlap correction
factors turn out to have a very mild generational dependence
\cite{A4CPV}, we can take them to be degenerate and parameterize
their effect by an overall multiplicative factor, $B_P^d$.

To improve on the mass insertion approximation we diagonalize the
one generation KK-zero mass matrix in Eq.~(\ref{M4KK}) to directly
obtain the physical coupling between a zero mode and the three KK
modes of the same generation, including the previously unaccounted
$(--)$ KK modes, while generational mixing effects are still
estimated using the spurion structure of Eq.~(\ref{C7Doverlap})
with all overlap matrices set to $\mathbbm{1}$
\cite{Azatov,IsidoriPLB,A4CPV}.

\begin{figure}[t]
\centering
\includegraphics[width=10 truecm]{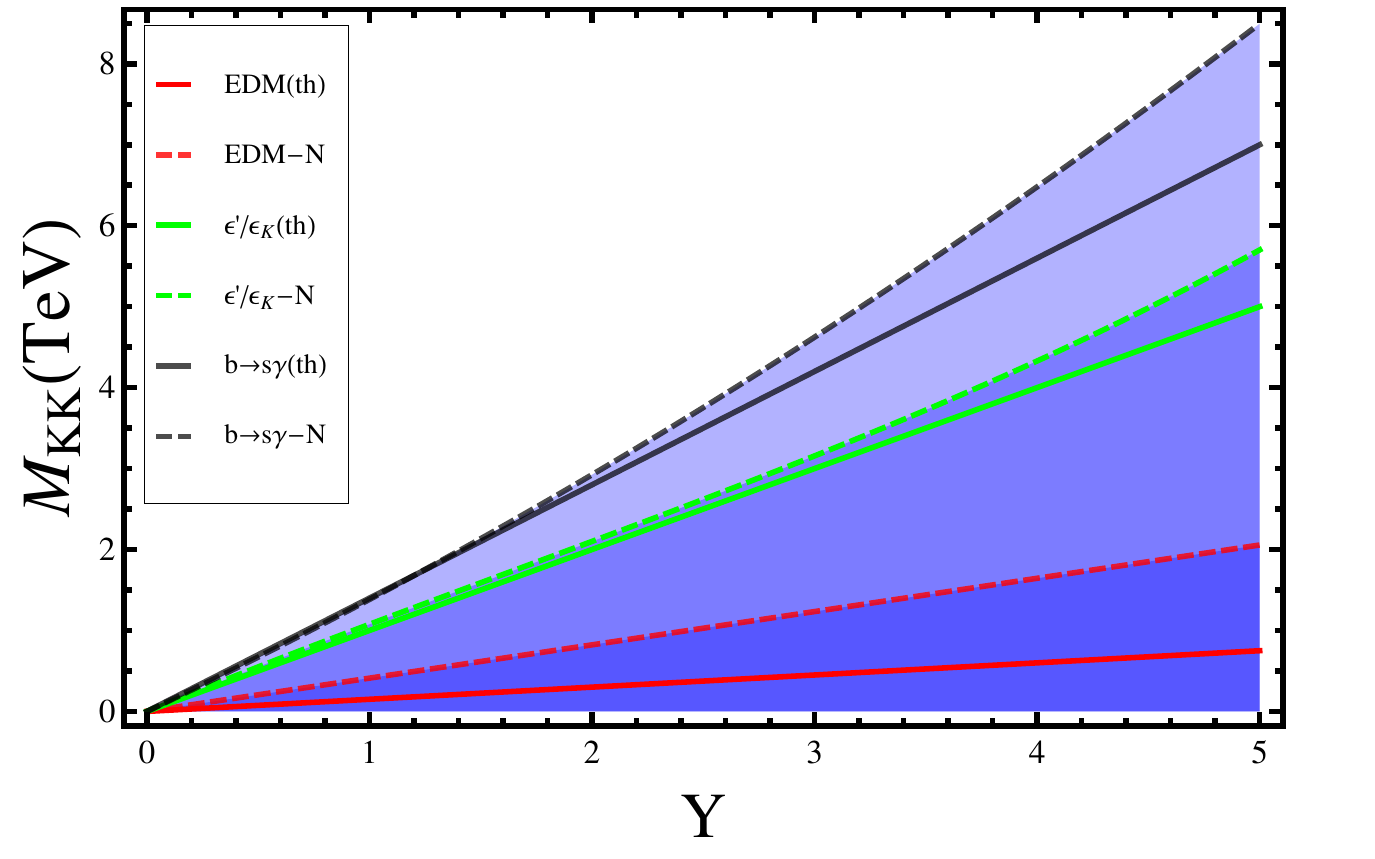}
\caption{ Bounds on the KK mass scale $M_{KK}$ as a function of
the overall Yukawa scale $Y$ for  the neutron EDM (bottom, red)
$\epsilon'/\epsilon$
 (center, green) and $b \rightarrow s\gamma$ (top, black). The analytical
(th) results (solid lines) are obtained within the one-generation
approximation combined with the spurion-overlap analysis (with the
exception of  the EDM (see text)) and compared with the numerical
(N) results (dashed lines) of the three-generation case. In both
cases, predictions are obtained for the model parameters that lead
to a realistic CKM matrix.}\label{PlotNumerical}
\end{figure}

The  down-type multiplicative overlap correction factor,
$B_P^{d}$, corresponding to the process in Fig.~\ref{fcncLoop},
will thus be extracted from
\begin{equation}
(\mathcal{A}_{ij})^{overlap}_D=\left. \frac{ \sum_n(
(O_L^{(d_i)_{KK}})^\dagger\hat{Y}_{d_i}^{KK}
 O_R^{(d_i)_{KK}})_{1n}
( (O_L^{(d_i)_{KK}})^\dagger\hat{Y}_{d_j}^{KK} O_R^{(d_i)_{KK}}
)_{n1}   } {\left (  M_{KK}^{d_i\, (n)}/M_{KK}   \right )}
\right|_{overlap}\, .\label{KK1genAppD}
\end{equation}
We use the above procedure for obtaining the constraints on the KK
scale arising from ${\mbox Re}(\epsilon'/\epsilon_K)$ and
$b\rightarrow s\gamma$. However, by using the same procedure, the
contribution to the neutron EDM turns out to be vanishing to first
order in $f_\chi^{u_i,d_i}(\tilde{x}_i^{u,d},\tilde{y}_i^{u,d})$.
We can  partially account for the latter by using the analog of
Eq.~(\ref{C7Doverlap}) for the up type RS-A$_4$ neutron EDM
contribution with non degenerate overlap correction factors. This
yields an $\mathcal{O}(f_\chi^2\Delta r)$ suppressed contribution,
where $\Delta r$ is the characteristic difference between nearly
degenerate overlap correction factors and is numerically of
$\mathcal{O}(0.01)$. The analytical estimations are then compared
with the numerical solution of the full three generations case,
where a diagonalization of the full $12\times 12$ matrix is
performed. The results of this comparison are reported in
Fig.~\ref{PlotNumerical}.

\subsection{The custodial anarchic case}
As a byproduct of our analysis we obtain the one generation KK
estimations for a flavor anarchic setup with an additional
$SU(2)_R$ doublet per generation and compare with the results of
\cite{Azatov,IsidoriPLB}. The anarchic mass matrix is obtained
from the matrix in Eq.~(\ref{M4KK}) by eliminating all Yukawas and
redefining $x=vY/M_{KK}$, where $Y$ is the characteristic
magnitude of the 5D anarchic Yukawa couplings. Generational mixing
effects will simply contribute $f_{Q_i}^{-1}f_{d_j}^{-1}$ for the
$(ij)$ component of the dipole operators in Eq.~(\ref{SM-O7}).
Together with the diagonalization of the mass matrix we obtain:
\begin{equation}
(C_{7\gamma(8g)}^d)_{ij}^{Cu.-An.}=\frac{f_{Q_i}^{-1}f_{d_j}^{-1}Y^3}{64\pi^2M_{KK}^2}
\left(r_{01}r_{10}(r_{11}+r_{111}+9r_{22}+r_{222})+r_{01}r_{101}(r_{11}+r_{111}+r_{22}+9r_{222})\right)
\label{AnarchyCust.}\end{equation} Using the anarchic bulk mass
assignments \cite{Agashe:2004cp} and $\beta_H=0$, which yields the
weakest bounds as in \cite{IsidoriPLB}, we obtain the constraints
on the KK mass scale from ${\mbox Re}(\epsilon'/\epsilon_K)$ and
$b\rightarrow s\gamma$ in the custodial anarchic case
\begin{equation} (M_{KK})_{\epsilon'/\epsilon}\gtrsim 1.4\,{\mbox
TeV} \qquad (M_{KK})_{b\rightarrow s\gamma}\gtrsim 0.5\,{\mbox
TeV}
\end{equation}

 \noindent When combined with the constraint
from $\epsilon_K$ \cite{IsidoriPLB}, which remains unchanged in
the custodial anarchic case the resulting bound on the KK scale
is: \begin{equation}(M_{KK})^{Cu.-An.}\gtrsim 6.6\,{\mbox TeV}
\end{equation} pushing the KK mass scale close to the limit of the
LHC reach.

\section{Conclusions}
We have illustrated a model based on an A$_4$ flavor symmetry and
implemented in a warped extra dimensional setup. The warped
geometry supplements us, as usual, with solutions to both the
gauge hierarchy problem and the fermion mass hierarchy.
Simultaneously, the bulk A$_4$ flavor symmetry and its distinct
SSB patterns towards the UV/IR branes by the A$_4$ flavons, $\Phi$
and $\chi$, allows us to obtain a TBM neutrino mixing pattern and
zero quark mixing as LO results, while NLO corrections induce
realistic quark mixing and small deviations from TBM. While the
experimentally allowed ranges for neutrino mixing angles
\cite{NuFogli} can also be explained with a bi-maximal mixing
pattern at LO with large NLO corrections \cite{Luca}, they can be
still accounted for when considering  NLO "cross-talk" and
"cross-brane" interactions in the RS-A$_4$ model. In the quark
sector, an exact matching of the CKM matrix is possible at the
price of increased tuning of NLO 5D Yukawa couplings. \\
\noindent The most stringent constraint on the RS-A$_4$ model
comes from the EWPM of the effective $Zb\bar{b}$ coupling.
Together with the perturbativity bound of the 5D top Yukawa
coupling it implies a constraint $M_{KK}\gtrsim4\,$TeV for
$m_h\approx 200\,$GeV and strongly constrains the common LH bulk
mass parameter. On the other hand the constraints arising from the
neutron EDM, $\epsilon'/\epsilon_K$ and $b\rightarrow s\gamma$
were shown both analytically and numerically (see
Fig.~\ref{PlotNumerical}) to induce milder constraints, dominated
by the one arising from $b\rightarrow s\gamma$.

\noindent To release the strong constraint from $Zb\bar{b}$ it is
relevant to consider the implications of embedding RS-A$_4$ in a
$P_{LR}$ extended custodial setup
\cite{AgashePLR,Casagrande_ECust}, where the dominant
contributions to the anomalous $Zb\bar{b}$ coupling are strongly
suppressed. However, since this embedding requires an extended non
trivial fermion sector, we expect it to increase the bounds from
the various FCNC processes we considered due to the presence  of
new KK modes and  a richer flavor structure. In addition, we
anticipate the typical prospects for observing $t\to cZ$ and $t\to
ch$ and the reduction of Higgs production cross sections
 \cite{Casagrande_ECust} to be mildly modified. A detailed
study of the various phenomenological implications of a $P_{LR}$
custodial RS-A$_4$ setup will be the subject of a future
publication.

\ack We thank the organizers of \emph{DISCRETE'10 - Symposium on
Prospects in the Physics of Discrete Symmetries} for  the
opportunity to present our work and for the lovely hospitality in
Rome.

\end{document}